# Influence of Trapping on the Exciton Dynamics of $Al_xGa_{1-x}As$ Films

A. Amo[1], M. D. Martin[1], Ł. Kłopotowski[1], L. Viña[1], A. I. Toropov[2] and K. S. Zhuravlev[2]

[1] *Dept. Física de Materiales, Universidad Autónoma de Madrid, E-28049 Madrid, Spain*
[2] *Institute of Semiconductor Physics, Pr. Lavrentieva, 13, 630090 Novosibirsk, Russia*

We present a systematic study on the exciton relaxation in high purity AlGaAs epilayers. The time for the excitonic photoluminescence to reach its maximum intensity (tmax) shows a non-monotonic dependence on excitation density which is attributed to a competition between exciton localization and carrier-carrier scattering. A phenomenological four level model fully describes the influence of exciton localization on tmax. This localization effect is enhanced by the increase of the Al content in the alloy and disappears when localization is hindered by rising the lattice temperature above the exciton trapping energy.

The simultaneous availability of commercial pulsed lasers and of nanostructurated semiconductors in the late 70s' concentrated the attention of the optical studies on carrier dynamics on these heterostructures, diverting it from bulk III-V materials. GaAs and AlGaAs are the constituent materials of many heterostructures like VCSELs, superlattices, microcavities, cavity LEDs or GRINSCH. The light emission characteristics of these heterostructures are fairly well understood as they have been the subject of intense study for the past decade.[1] A precise knowledge of the properties of their elemental constituents is needed for the complete understanding of the behaviour of these devices and for the design of future systems based on these materials. However, little attention has been paid to the luminescence dynamics of bulk GaAs, and even less to that of AlGaAs. Only recently some works,[2-4] which complement but not complete older scattered studies,[5-8] have investigated the emission dynamics just in GaAs, since good alloy samples have not been available. Yet, important issues in III-V alloys, such as the influence of defects or Aluminium content on the exciton dynamics, have not been addressed.

In this letter, we present a systematic study of the effect of exciton localization on the photoluminescence (PL) build up in bulk GaAs and $Al_xGa_{1-x}As$ with x up to 0.05. In the past, exciton localization effects in bulk GaAs[7,8] and quantum wells (QWs),[9-11] have mainly been studied through their influence on the excitonic luminescence decay time. Still we have found that exciton localization has a stronger influence on the rise dynamics, which we characterize by the time the free-exciton PL takes to reach its maximum, $t_{max}$.

We investigated four high quality $Al_xGa_{1-x}As$ epilayers of 2.5 μm thickness, with Al concentration x = 0, 0.015, 0.03 and 0.05, grown by Molecular Beam Epitaxy.[12] All the samples, which were nominally undoped, showed p-type conductivity with hole concentrations in the range $1-8 \times 10^{14}$ cm$^{-3}$. The samples were mounted on a cold finger cryostat, which enabled precise control of the temperature between 5-45 K, and were non-resonantly photoexcited with a Ti:Sapphire laser that produced 2 ps long pulses. The PL was energy- and time-resolved by a synchroscan streak camera in conjunction with a spectrometer. The time and energy resolution of the overall setup is better than 12 ps and 0.4 meV, respectively. A low-temperature cw PL excitation characterization of the samples (not shown here), allowed a precise determination of the free and bound exciton emission peaks (FX and BXs, respectively). Despite the presence of localization centres, the narrowness of the excitonic lines (full width at half maximum below 1 meV for the GaAs epilayer) guarantees the good quality of the samples.

Figure 1 shows the PL spectra at 5 K and low excitation density ($n = 1.8 \times 10^{14}$ cm$^{-3}$) of the GaAs (black line) and x = 0.03 (grey line) samples 1 ns after the excitation. The PL is dominated by the emission from FX and BXs – $A^0$-X, $D^0$-X – as well as by electron-acceptor (e-$A^0$) recombination (peaks have been identified following Refs. 7, 13). An exciton localization energy in acceptor related sites, of about 2.7 meV, can be extracted from the peaks' position. At low excitation densities, the BX luminescence overcomes that of the FX as the Al fraction in the alloy is augmented, due to the increase in the number of defects. However, at high excitation densities ($n \sim 7.5 \times 10^{15}$ cm$^{-3}$) the PL is completely dominated by the FX emission in all samples. In the analysis of the time-resolved PL we will concentrate on the FX and $A^0$-X lines.

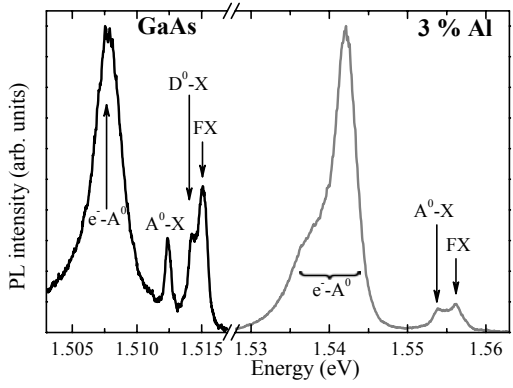

FIG. 1. Low temperature (5 K) PL spectra of the GaAs (black line) and x = 0.03 (grey line) samples recorded at a delay of 1 ns after excitation (excitation density $1.8\times10^{14}$ cm$^{-3}$; excitation energy 1.630 eV).

Figure 2 depicts the time evolution of the FX (solid points) and $A^0$-X (BX; open points) emissions for the GaAs and the x = 0.03 samples under the same conditions as in Fig. 1. $t_{max}$, which can be easily assigned, as indicated in the figure by the horizontal bars, is considerably longer for the GaAs than for the x = 0.03 sample. We have observed that $t_{max}$ is also longer for the BXs than for the FXs (Fig. 2) in all investigated samples, revealing longer energy relaxation processes in the case of the BXs. In addition, as can be clearly seen for GaAs in Fig. 2(a), the BX presents a concave time evolution at short times after the excitation, while that of the FX is convex. The concave curvature of the BX traces indicates that the build up of the BX population results from a multi-step relaxation process. Both facts, the longer $t_{max}$ for FX than for the BX and the concave curvature, evidence that the source of the BX luminescence is the trapping of FX with centre of mass momentum K~0, in a cascade process similar to that observed when carriers from the barrier are trapped in quantum dots.[3, 14]

In order to account for this cascade process we have considered a four-level rate-equation model as depicted in the inset of Fig. 2(b). The excitation pulse creates electron-hole pairs which rapidly (< 20 ps)[15] form excitons with large K (level 3 in the inset). These FXs relax their kinetic energy towards the radiative K~0 states (level 2) via emission of acoustic phonons,[2] with a characteristic time $\tau_k$. FXs can then either radiatively recombine ($\tau_r$) or get trapped in localization sites ($\tau_B$) giving rise to a BX population (level 1), which can also radiatively recombine ($\tau_{rB}$). The differential rate equations that describe the dynamics of such a four-level model can be easily analytically integrated.

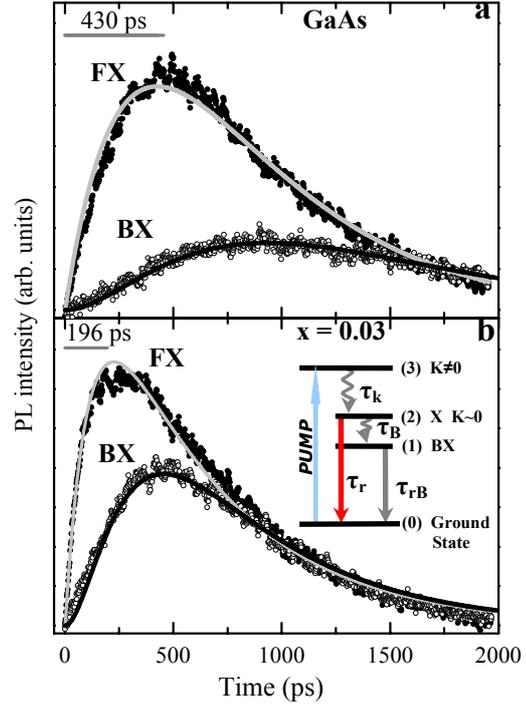

FIG. 2. (a) Time evolution of the FX (solid dots) and the BX ($A^0$-X; open dots) for the GaAs sample (excitation density $1.8\times10^{14}$ cm$^{-3}$; T= 5 K). (b) Same as (a) for the Al$_x$Ga$_{1-x}$As x = 0.03 sample. $t_{max}$ is indicated by horizontal bars. The solid lines are fits to the model described in the text ($\tau_k$ = 464 ps, $\tau_B$ = 1240 ps and $\tau_r = \tau_{rB}$ = 565 ps for GaAs). The inset shows the level scheme and the transitions considered in the model.

Figure 2 also shows the *simultaneous* least squares fits of the temporal traces extracted from the four-level model to the FX and BX data. For the sake of reducing the number of fitting parameters, the fits were performed assuming the same recombination times for FX and BX ($\tau_r = \tau_{rB}$). Good agreement is found between the experiments and the fits. From the fitting parameters we have found a trapping time $\tau_B$ four times shorter in the x = 0.03 epilayer than in the GaAs one, evidencing a density of traps about four times greater in the x = 0.03 sample ($\tau_B \propto$[traps]$^{-1}$).[7]

Figure 3(a) shows the dependence of the FX-$t_{max}$ on excitation density at 5 K for the four investigated epilayers. All the samples show a similar non-monotonic behaviour,[16] with a maximum $t_{max}$ at a carrier density $n\sim1.2\times10^{16}$ cm$^{-3}$ $\equiv n_{x\text{-}x}$. This density corresponds to a mean distance between excitons of ~40 nm, which is of the same order of magnitude as the exciton Bohr radius ($a_B$) in GaAs (11.2 nm). Hence, for $n > n_{x\text{-}x}$, exciton-exciton elastic scattering is an important source of scattering events and results in a fast relaxation of FX with large K toward the radiative states with K~0, as previously reported in QWs.[15, 17] Exciton-free carrier collisions also take place for the highest excitation densities considered in this work, and take part in the fast relaxation of carriers as also demonstrated in the case of QWs.[17] The higher the carrier density in the sample the more efficient these processes become, resulting in a decrease of $t_{max}$ with increasing excitation density. The threshold density $n_{x-x}$ only depends on $a_B$ and thus is expected to be independent of the Al content for the low concentrations considered in this letter, as borne out by our experiments.

For $n < n_{x-x}$, FX-$t_{max}$ increases with increasing excitation density. This behaviour can be understood if trapping of FX in localized states is taken into account in the framework of the proposed four-level model. The PL spectra in all investigated samples are dominated by BX emission for densities below $n_{x-x}$. Therefore, at the lowest studied densities, there are two mechanisms for the depletion of excitons from the K~0 FX level: (i) exciton radiative recombination, and (ii) trapping of FX with K~0 into localized BX states. As the excitation density is increased, BX trapping states are gradually filled up to their saturation. When $n$ is high enough, the number of available localization sites is small compared to the FX population, and the trapping channel (ii) has very little effect on the dynamics of the FX population. The direct consequence of the saturation of trapping centres, and therefore, the closing of one of the depleting channels, is the increase of $t_{max}$. This situation is reproduced by the model, and its extreme case corresponds to the elimination of the BX level: fixing the values of $\tau_k$ and $\tau_r$ obtained from the fits of Fig. 2(a), and taking $\tau_B \rightarrow \infty$, the model would give a FX temporal trace with $t_{max}$ considerably longer –512 ps– than that obtained in the presence of the localization channel –430 ps–. This is in qualitative agreement with the increase of $t_{max}$ with carrier density shown in Fig. 3(a).

The increase of the Al content in the alloy results in two effects: (i) an acceleration of the dynamics due to the enhancement of alloy scattering, which results in the reduction of $t_{max}$ for any $n$; (ii) a larger number of alloy defect-related traps. As already mentioned above, a higher density of localization centres produces a reduction of the trapping time $\tau_B$ which, according to the four level model, results in an additional decrease of $t_{max}$ for $n < n_{x-x}$.

In order to gain insight into the influence of trapping on the FX dynamics, time resolved experiments have been performed at different lattice temperatures. The results for GaAs are shown in Fig. 3(b), which depicts $t_{max}$ versus excitation density for temperatures up to 45 K. For $T \leq 30$ K, the observed behaviour follows the trends discussed above for low T (non-monotonous dependence of $t_{max}$ on $n$). In contrast, if the lattice temperature is raised above the exciton localization energy (2.7 meV ↔ 31.3 K), the BXs are ionized and trapping is hindered. In the absence of the trapping depletion channel, according to the model no dependence of $t_{max}$ on excitation density for $n < n_{x-x}$ is expected, as found experimentally for T = 38 K and T = 45 K (Fig. 3b).

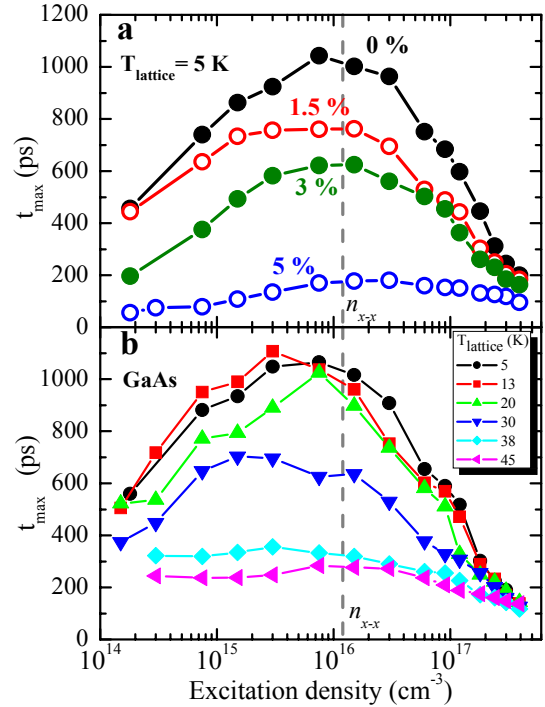

FIG. 3. (a) Time for the free exciton to reach its maximum intensity, $t_{max}$, as a function of excitation density in the four investigated $Al_xGa_{1-x}As$ epilayers at a lattice temperature of 5 K. Aluminium content is indicated in each curve. (b) $t_{max}$ as a function of excitation density for the GaAs sample at different lattice temperatures.

In conclusion, we have presented a systematic study of the effect of exciton localization on the PL rise time in AlGaAs. For high excitation densities, exciton-exciton and exciton-carrier scattering dominate the dynamics. For low excitation densities, $t_{max}$ is strongly influenced by the trapping of FX into BX states. The competition between localization and carrier-carrier scattering obtains a non-monotonic dependence of the rise time, $t_{max}$, on excitation power. A four-level phenomenological model fully describes the excitation-density, Aluminum-content and lattice-temperature dependence of the rise time.

This work was partially supported by the Spanish MCYT (MAT2002-00139), the CAM (07N/0042/2002), the network "Marie-Curie: Physics of Microcavities" (MRTN-CT-2003-503677) and the Russian Basic Research Fund (grant no. 04-02-16774a). A. A. acknowledges a scholarship of Spanish Secretaría de Estado de Educación y Universidades (MEC).